\documentclass[a4paper,superscriptaddress,showpacs,twocolumn]{revtex4}
\usepackage{graphicx}
\usepackage{amsmath}
\usepackage{amsfonts}

\def\vek#1{\boldsymbol{#1}}

\def\mat#1{\boldsymbol{\mathsf{#1}}} 
\def\vvek#1{\mathbb{#1}}
\def\mmat#1{\mathbb{#1}} 
\def\a{\tilde{\mu}} 

\def\up{\uparrow}
\def\dw{\downarrow}
\def\updw{{\up/\dw}} 
\def\int{\mathrm{int}}
\def\sp{\mathrm{sp}}
\def\ch{\mathrm{ch}}
\def\co{\mathrm{Co}}
\def\cu{\mathrm{Cu}}

\def\tabI{%
\begin{table*}
\begin{ruledtabular}
\begin{tabular}{lccc}
 & $j_\sp$ [$10^9$A/m$^2$] &  
  $\Delta\mu$ [meV]  & MR [\%]
\\ \hline \hline
(Cu/Co)$^2$ $\up\up$  S-constant ($+$) &
14.85 [14.93] &  -0.223 [-0.220] & 0.483 [0.485]
\\ \hline
(Cu/Co)$^2$ $\up\up$ column ($\square$) &
19.0 [19.8] & -0.267 [-0.276] & 0.63 [0.66]
\\ \hline
(Cu/Co)$^2$ $\up\up$ constriction ({\large $\diamond$}) &
29.4 [33.7] & -0.052 [-0.009] & 1.01  [1.17]
\\ \hline
(Cu/Co)$^2$ $\up\up$ constr, Co1 infinite & 
28.7 [23.3] & -0.050 [0.001] & 1.43 [2.48]
\\ \hline \hline
(Cu/Co)$^3$ $\up\up\up$ S-constant ($\times$) &
19.51 [19.67] & -0.078 [-0.077] & 0.444 [0.448]
\\ \hline 
(Cu/Co)$^3$ $\up\up\up$ column  ($\bigtriangleup$) &
23.6 [24.5] & -0.171 [-0.184] & 0.55 [0.57]
\\ \hline 
(Cu/Co)$^3$ $\up\up\up$ constriction ($\bigtriangledown$) &
31.0 [33.9] & -0.019 [-0.005] & 0.74 [0.81]
\\ \hline \hline
(Cu/Co)$^3$ $\up\up\dw$ S-constant  ({\large $\ast$}) &
5.94 [5.93] & -0.442 [-0.438] & 0.116 [0.115]
\\ \hline 
(Cu/Co)$^3$ $\up\up\dw$ column ({\large $\triangleright$}) &
8.08 [8.49] & -0.490 [-0.494] & 0.162 [0.171]
\\ \hline 
(Cu/Co)$^3$ $\up\up\dw$ constriction ({\large $\triangleleft$}) &
9.7 [10.5] & -0.458 [-0.456] & 0.192 [0.210]
\end{tabular}
\end{ruledtabular}
\caption{Values of $j_\sp$, $\Delta\mu=\mu_\up-\mu_\dw$ at the
  position of free Co2 layer for (Cu/Co)$^2$ and (Cu/Co)$^3$
  structures. Magnetoresistance ration (MR) is determined 
  between first and last Co/Cu interface. All values are determined as
  averadge over whole pillar cross-section area. In square
  brackets we present values calculated from modified 1D VF
  formalism, which takes into account a~variable cross-sectional area of
  the layers 
  (Sect.~\ref{s:cr1dm}). Symbols in parentheses denote structure
  notation in Fig.~\ref{f:jR}.}
\label{t:tab}
\end{table*}
}

\def\figI{%
\begin{figure}
\begin{tabular}{rl}
\kern0mm\lower-9mm\hbox{(a)}&
\hspace*{9.5mm}\includegraphics[scale=0.7]{Rce.1}\\[-2mm]
\kern0mm\lower-9mm\hbox{(b)}&
\includegraphics[scale=0.7]{Roe.1}\\[-2mm]
\kern0mm\lower-9mm\hbox{(c)}&
\includegraphics[scale=0.7]{Rml.1}
\end{tabular}
\caption{A~general form of (a) close-end SDRE (b) open-end
  SDRE (c) classical multilayer structure having both ends
  open-ended. The arrow shows  positive both structure and current
  direction. In all cases, SDRE has three layers, $M=3$.}
\label{f:sdre}
\end{figure}
}

\def\figII{%
\begin{figure*}
\includegraphics[scale=0.8]{exam.1}
\caption{The schema of a~circuit example, consisting of three node, two
open-end SDRE and three close-end SDRE. The arrows
parallel to SDRE shows the defined direction of SDRE (and hence
also positive current direction).}
\label{f:netsch}
\end{figure*}
}

\def\figIII{%
\begin{figure}
\setlength{\unitlength}{0.01\textwidth}
\includegraphics[width=0.38\textwidth]{res.1}
\put(-6,24){\includegraphics[width=0.09\textwidth]{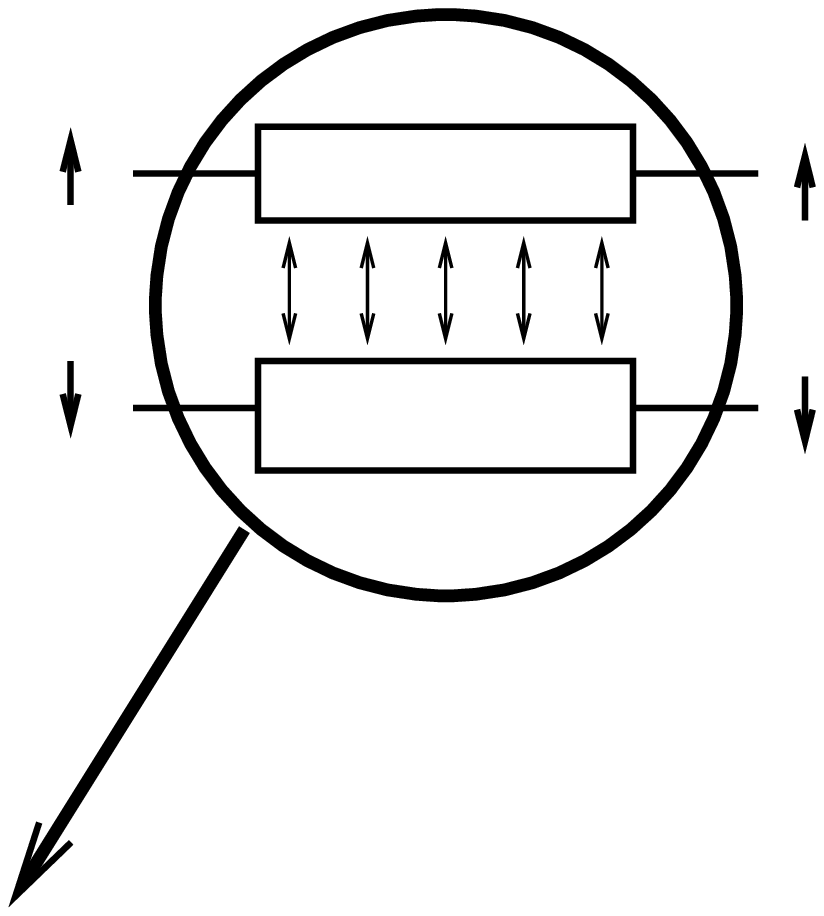}}
\caption{An example of dividing nanostructure into 3D circuit of SDRE. Each
  wire represents a ``bus'' of channel-up and channel-down. The inset
  remark that each SDRE consists of resistivity for channel-up, for
  channel-down and a spin-flip-scattering resistance between both
  channels. The large resistors denotes for interface resistivity
  [Eq.~(\ref{eq:Kint})].} 
\label{f:grid}
\end{figure}
}

\def\figIV{%
\begin{figure}
\includegraphics{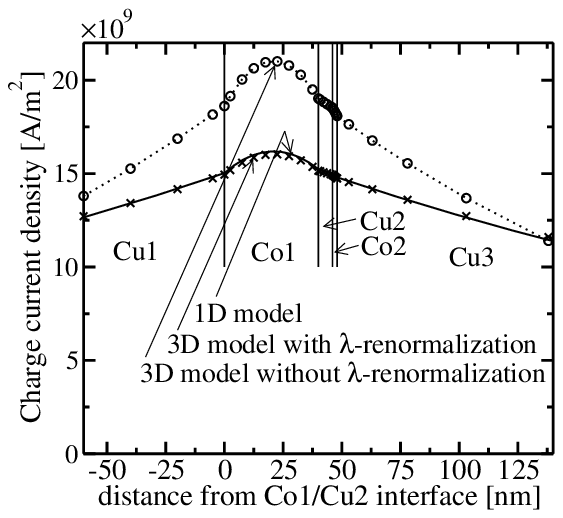}
\caption{j$_\sp$ through (Cu/Co)$^2$ structure with constant
  cross-sectional area $S$ calculated by 1D model (line), by our 3D
  model with ($\times$) and without ($\circ$)
  $\lambda$-renormalization.}
\label{f:2d3d}
\end{figure}
}

\def\figV{%
\begin{figure*}
\includegraphics{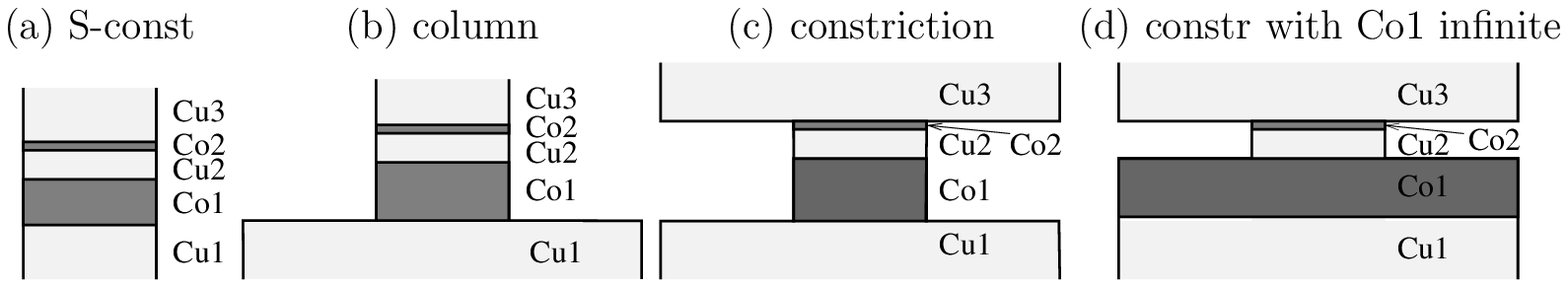}
\caption{Sketches of studied type structures. (a) S-constant
  (infinitely long nanowire with constant cross-section area $S$), (b)
  column (infinitely long pillar deposited on infinitely large Cu1
  layers) and (c) constriction (both Cu3 cover and Cu1 buffer layers
  are infinitely large) (d) constriction with Co1 infinite layer.}
\label{f:types}
\end{figure*}
}

\def\figVI{%
\begin{figure*}[floatfix]
\includegraphics[scale=0.95]{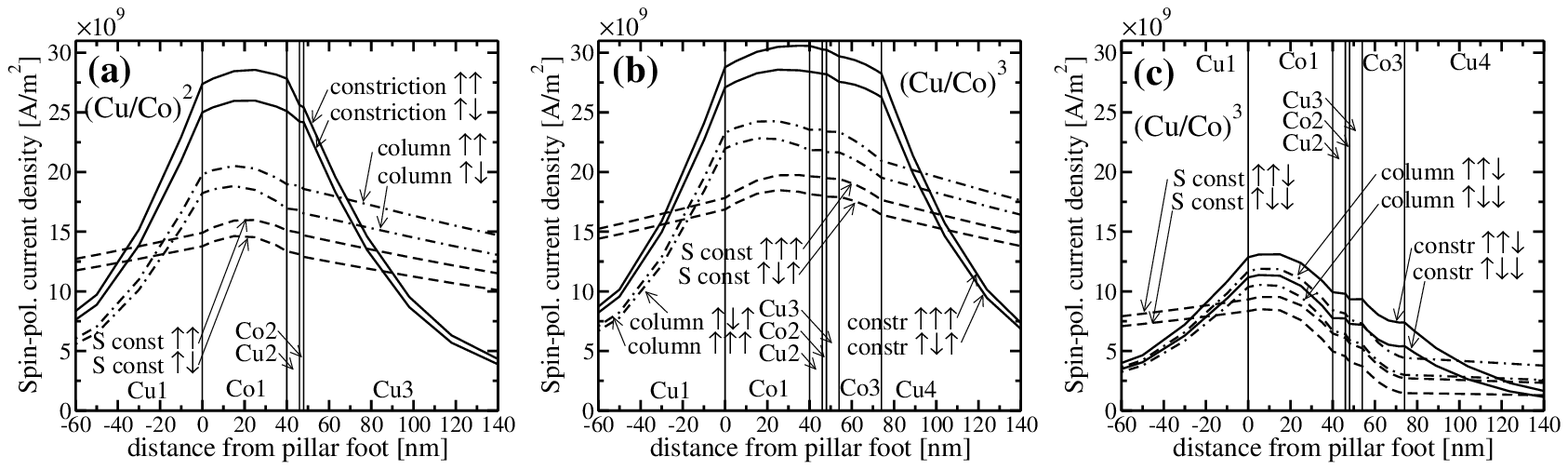}
\caption{$j_\sp$ along center axis
  of the (a) (Co/Cu)$^2$ $\up\up$, $\up\dw$ (b) (Co/Cu)$^3$
  $\up\up\up$, $\up\dw\up$ (c) (Co/Cu)$^3$ $\up\up\dw$, $\up\dw\dw$
  for S-constant, column and constriction structure types
  [Fig.~\ref{f:types}].}
\label{f:crsp}
\end{figure*}
}

\def\figVII{%
\begin{figure}[b]
\includegraphics{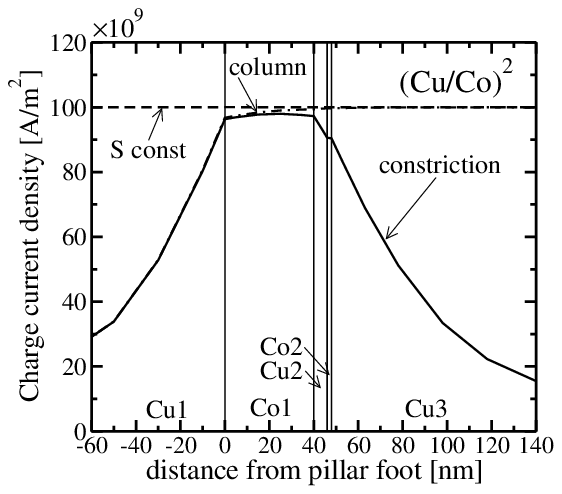}
\caption{$j_\ch$ along center axis of the
  (Co/Cu)$^2$ 
  structure for S-constant, column and constriction structure types 
  [Fig.~\ref{f:types}].} 
\label{f:crch}
\end{figure}
}

\def\figVIII{%
\begin{figure*}
\includegraphics{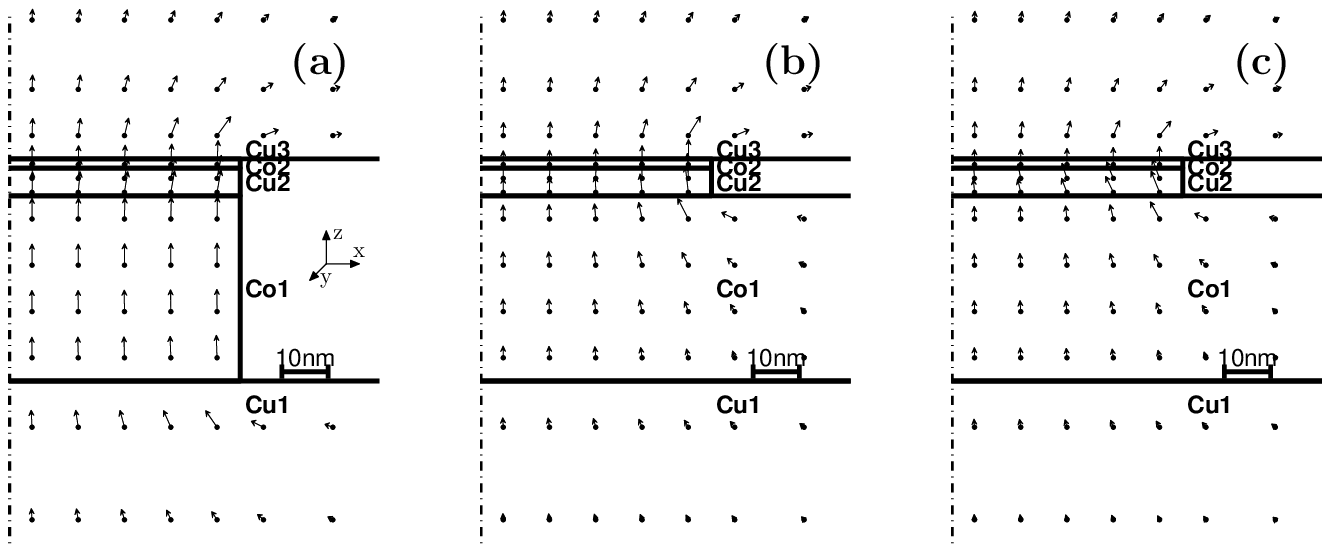}
\caption{Spin polarized current in (Cu/Co)$^2$ 
  constriction-type structure (a) without infinitely large Co1 layer,
  $\up\up$ (b) with infinitely large Co1 layer $\up\up$ (c) as (b) but
  $\up\dw$.}
\label{f:pl2}
\end{figure*}
}

\def\figIX{%
\begin{figure*}
\includegraphics{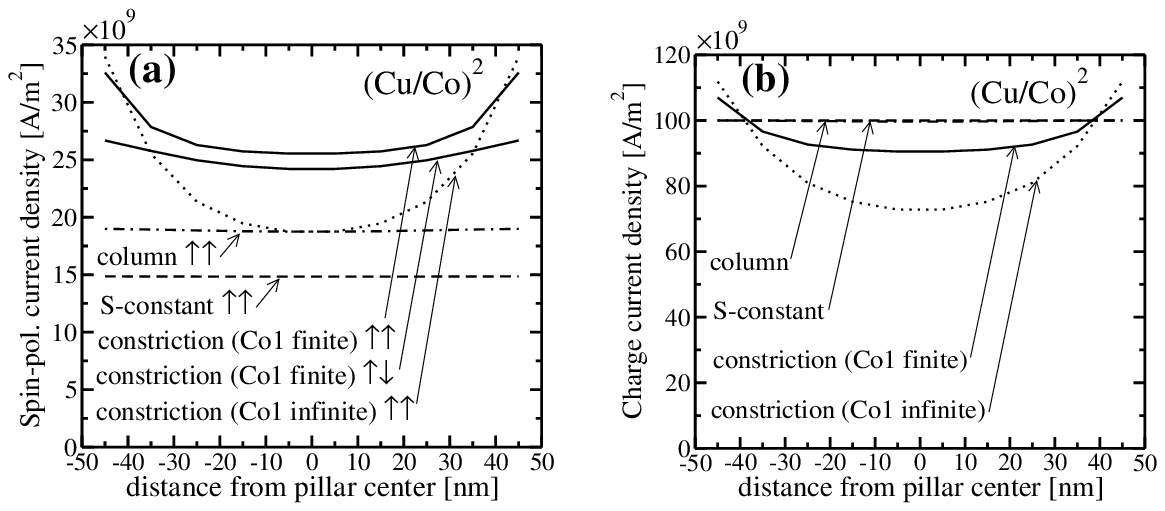}
\caption{(a) $j_\sp$ and (b) $j_\ch$
   in the middle of the Co2 layer inside (Cu/Co)$^2$ structures.}  
\label{f:prftra}
\end{figure*}
}

\def\figX{%
\begin{figure*}
\includegraphics{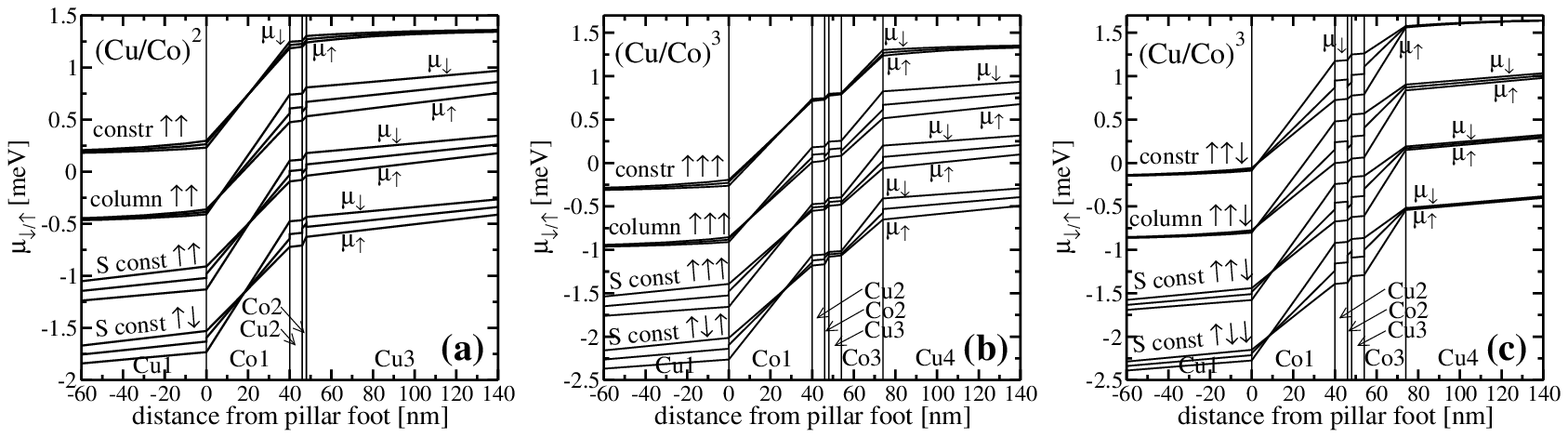}
\caption{Electro-chemical potential $\mu_{\up/\dw}$ along center
  axis of the (a) (Cu/Co)$^2$ $\up\up$, $\up\dw$ (b) (Cu/Co)$^3$
  $\up\up\up$, $\up\dw\up$ and (c) (Cu/Co)$^3$ $\up\up\dw$,
  $\up\dw\dw$ for S constant, column and constriction structure types
  [Fig.~\ref{f:types}].}
\label{f:mu}
\end{figure*}
}

\def\figXI{%
\begin{figure*}
\includegraphics{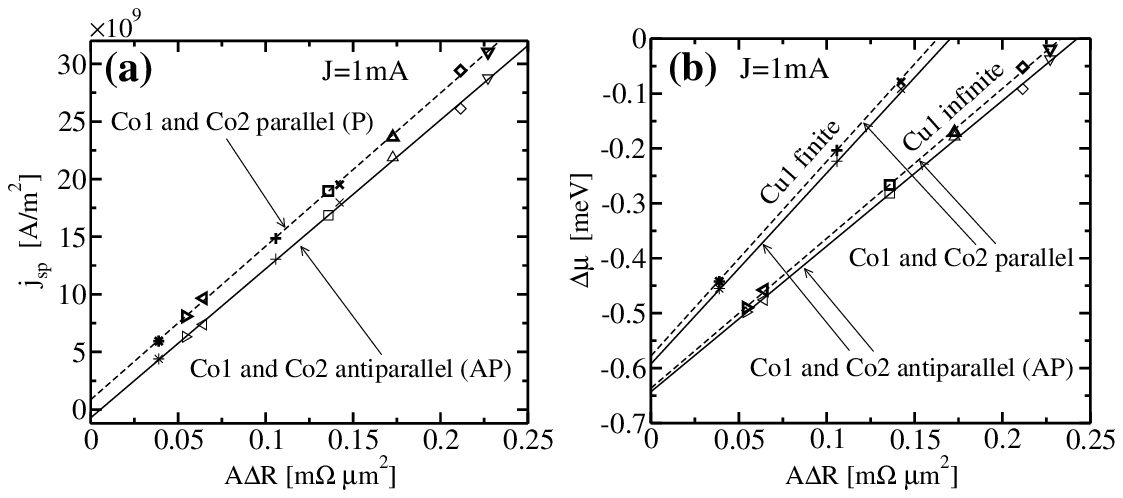}
\caption{Dependence of (a) $j_\sp$ and (b) $\Delta\mu$ on $A\Delta
  R$ for different studied (Cu/Co) structure, symbol notation in
  Tab~\ref{t:tab}. Lines are the best linear
  fits, dashed line and bold symbols (full line and normal symbols)
  denote parallel (antiparallel) Co1 and Co2 layers. Cu1 infinite
  means column and constriction types, Cu1 finite means S constant
  type.}
\label{f:jR}
\end{figure*}
}

\def\figXII{%
\begin{figure*}
\includegraphics{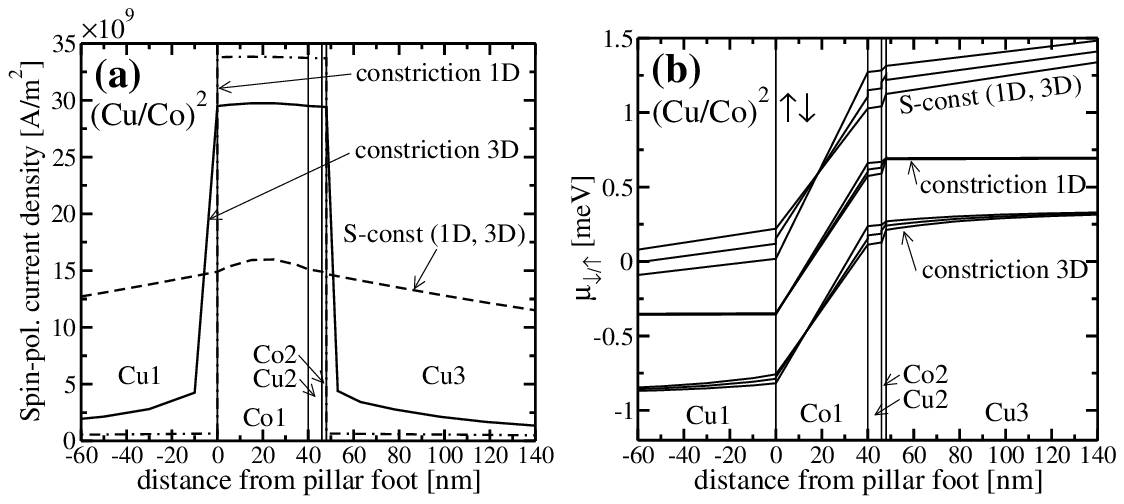}
\caption{(a) averaged value of spin-polarized current and (b)
  $\mu_\updw$ in the structure center calculated for constriction
  type structure with and without Co1 infinite layer in the
  (Cu/Co)$^2$ $\up\dw$ compared with 1D
  calculations taking into account infinite Cu layers. See
  Section~\ref{s:cr1dm} for detail.}
\label{f:1dm}
\end{figure*}
}

\begin{document}

\title{3-dimensional distribution of spin-polarized current:
  application to (Cu/Co) pillar structures}

\author{J. Hamrle}
\affiliation{
FRS, The Institute of Physical and Chemical Research (RIKEN), 
2-1 Hirosawa, Wako, Saitama 351-0198, Japan}
\affiliation{CREST, Japan Science \& Technology Corporation, Japan}

\author{T. Kimura}
\affiliation{
FRS, The Institute of Physical and Chemical Research (RIKEN), 
2-1 Hirosawa, Wako, Saitama 351-0198, Japan}
\affiliation{CREST, Japan Science \& Technology Corporation, Japan}

\author{T. Yang}
\affiliation{
FRS, The Institute of Physical and Chemical Research (RIKEN), 
2-1 Hirosawa, Wako, Saitama 351-0198, Japan}
\affiliation{CREST, Japan Science \& Technology Corporation, Japan}

\author{Y. Otani}
\affiliation{
FRS, The Institute of Physical and Chemical Research (RIKEN), 
2-1 Hirosawa, Wako, Saitama 351-0198, Japan}
\affiliation{CREST, Japan Science \& Technology Corporation, Japan}
\affiliation{ISSP, University of Tokyo, Kashiwa-shi, Chiba 277-8581, Japan}

\date\today

\begin{abstract}
  We present a formalism determining spin-polarized current and
  electrochemical potential inside arbitrary electric circuit within
  diffusive regime for parallel/antiparallel magnetic states. When
  arbitrary nano-structure is expressed by 3-dimensional (3D) electric
  circuit, we can determine 3D spin-polarized current and
  electrochemical potential distributions inside it. We apply this
  technique to (Cu/Co) pillar structures, where pillar is terminated
  either by infinitely large Cu layer, or by Cu wire with identical
  cross-sectional area as pillar itself. We found that infinitely
  large Cu layers work as a strong spin-scatterers, increasing
  magnitude of spin-polarized current inside the pillar twice and
  reducing spin accumulation nearly to zero.  As most experimentally
  studied pillar structures are terminated by such a infinitely large
  layers, we propose modification of standard Valet-Fert formalism to
  simply include influence of such infinitely large layers.

\end{abstract}


\maketitle



\section{Introduction}
\label{s:intro}

Spin injection, transport, and detection are key factors in the field
of magnetoelectronics. Especially, magnetization reversal using
spin-polarized current is of great interest
\cite{weg99,alb00,kat00,sun03,oez03,che04,fab03} due to its potential
technological applications such as MRAM \cite{eng02}, spin transistor
\cite{joh93} or spin battery \cite{bra02}.

To understand and optimize spin-transport behavior in such devices, it
is important to know a current distribution in it.  Particularly for
MRAM applications, we have to know spin-current magnitude and
distribution to optimize the current density necessary for
spin-injection induced magnetization reversal.  Variety of formalisms
calculating magnetoelectronic transport in one dimension (1D) even for
non-collinear magnetization has been proposed
\cite{val93,bra00,wai00,bau03,zha04}.

However, up to now, the spatial (3D) calculation of the spin-polarized
current have been missing. To obtain spatial distribution of
spin-polarized current (and spin accumulation) in a given structure,
we express such a structure as a~3D circuit of
spin-dependent-resistor-elements (SDRE), wherein the propagation is
regarded as 1D problem \cite{val93,jed03}.

Resistor circuit network has been already used in \cite{ern02} to
simulate current lines in metal/insulator multilayers.
However, in this case, they did not use spin-polarized
current.  Ichimura et al. \cite{ich04} have calculated 2D distribution
of spin polarized current for Co/Al lateral spin valve structure. 
They did not use the
electric circuit, but directly solved  Poison equation by means of
finite element method.
 
This article is organized as follows: Sections \ref{s:bases} and
\ref{s:int} present a~matrix approach to calculate 1D diffusive
propagation of spin-polarized current and potential in the single
SDRE.  Section~\ref{s:multilayer} shows how this formalism can be
applied to a~simple multilayer structure.  This approach is
just compact matrix rewriting the 1D Valet-Fert (VF) formalism
\cite{val93}. Section \ref{s:net} explains how to solve general
electric circuit and Section \ref{s:grid} tells how to divide
nanostructure into circuit of
SDREs. Finally, in Section~\ref{s:cuco} we apply our calculations to
(Cu/Co)$^2$, and (Cu/Co)$^3$ pillar structures, where cross-sectional
areas of the first and last layers are assumed either the same as 
the pillar, or infinitely large. We show how the presence of such infinitely
large layers influence currents and spin accumulation profiles. Finally
Section~\ref{s:cr1dm} shows how to modify VF formalism to describe
influence of the infinitely large layers.

\section{Diffusive transport regime}
\label{s:bases}

In the diffusive transport regime,  equations deriving the spatial
distribution of electrochemical potential $\mu_{\updw}$ and
spin-polarized current density $\vek{j}_{\updw}$ inside ferro or
non-magnetic metals are
\cite{son87,val93,jed03, tak03}
\begin{gather}
\label{eq:diffmu}
  \nabla^2(\mu_\up-\mu_\dw)=\frac{\mu_\up-\mu_\dw}{\lambda^2},
\\[1mm]
\label{eq:diffj}
  \vek{j}_{\up/\dw}=-\frac{\sigma_{\up/\dw}}{e}\nabla \mu_{\up/\dw},
\end{gather}%
where $\lambda$ is spin-flip diffusion length,
$\sigma_{\up/\dw}=\sigma(1\pm\beta)/2$ conductivities for up and down
channels, respectively and $e$ the electron charge assumed to be $e=1$
in this article.

\figI
Our model is based on the circuit of SDRE, consisting of layers and
interfaces (Fig.~\ref{f:sdre}).  We first express response of a~single
layer.  When Eqs.~(\ref{eq:diffmu})(\ref{eq:diffj}) are solved in 1D,
the profiles of $\mu_\up$ and $\mu_\dw$ (hereafter denoted as
$\mu_\updw$) inside a~given material along $x$-axis is given by
\cite{jed03,val93}
\begin{multline}
\label{eq:defmu}%
\hspace*{-3mm}
 \mu_{\up/\dw}(x+L)
       = \a(x)+\frac{j_\ch e}{\sigma} L
       \pm c(x) \frac{\sigma}{\sigma_{\up/\dw}} \exp[-L/\lambda]
\\
       \pm d(x) \frac{\sigma}{\sigma_{\up/\dw}} \exp[L/\lambda],
\end{multline}%
where $j_\ch=j_\up+j_\dw$ is charge current density and positive current
direction is towards positive $x$-direction. Energies $c(x)$ and $d(x)$ are
amplitudes of exponential dumping of $\mu_\updw$ and the energy $\a(x)$ is
the asymptotic electrochemical potential equivalent to weighted
average of electrochemical potentials $\a=(\sigma_\up\mu_\up+\sigma_\dw
\mu_\dw)/\sigma$.  The energies $\a$, $c$ and $d$ will be determined
later from boundary conditions.

The relation between $\mu_{\up/\dw}$, $j_{\up/\dw}$ and $\a$, $c$, $d$
 used in Eqs.~(\ref{eq:diffj}) and (\ref{eq:defmu})
can be expressed in a~compact matrix form
\begin{gather}
  \displaystyle
  \begin{bmatrix}
    \mu_\up \\ \mu_\dw \\ uJ_\up \\ uJ_\dw
  \end{bmatrix}_x  =
  \begin{bmatrix}
    1 & \frac{\sigma}{\sigma_\up} & \frac{\sigma}{\sigma_\up} & 0
    \\
    1 & -\frac{\sigma}{\sigma_\dw} & -\frac{\sigma}{\sigma_\dw} & 0    
    \\
    0 & -\frac{u\sigma S}{e \lambda} & \frac{u\sigma S}{e \lambda}
    & \frac{\sigma_\up}{\sigma}
    \\
    0 & \frac{u\sigma S}{e \lambda} & -\frac{u\sigma S}{e \lambda}
    & \frac{\sigma_\dw}{\sigma}    
  \end{bmatrix}
  \cdot
  \begin{bmatrix}
    \a \\ c \\ d \\ uJ_\ch
  \end{bmatrix}_x
\notag
\\[1mm]
  \vek{H}=\mat{D}\cdot\vek{F},
\label{eq:defD}
\end{gather}%
where $J_{\up/\dw}=Sj_{\up/\dw}$ is up/down spin-polarized current with 
cross-section area $S$ of a given layer, $J_\ch=J_\up+J_\dw$ is a~charge
current flowing through layer. The scaling factor $u$ which has no physical
meaning, is introduced to adjust units of newly defined $\vek{H}$
and $\vek{F}$ vectors. For numerical calculations, value of $u$ should
adjust order of $\mu_{\up/\dw}$ and $uJ_{\updw}$. When expressing
any measurable quantities (e.g.\ $\mu_\updw$, $J_\updw$, resitances),
they obviously do not depend on~$u$.

The left side of Eq.~(\ref{eq:defD}) ($\vek{H}$-vector) contains
variables, which are conserved across interfaces or electric nodes. On
the other hand, the right side of Eq.~(\ref{eq:defD})
($\vek{F}$-vector) contains variables, which are used to calculate
propagation of the $J_\updw$ and $\mu_\updw$ through layers
[Eq.~(\ref{eq:defmu})].  Hence, the dynamic $\mat{D}$-matrix relates
coefficients at the boundary to coefficients for propagation. This
approach is well-known in the optics of anisotropic media
\cite{yeh80,vis86}.

Propagation of $\vek{F}$-vector through layer with the thickness
(``length'') $L$ is expressed by a propagation $\mat{P}$-matrix
[Eq.(\ref{eq:defmu})]
\begin{gather}
  \displaystyle
  \begin{bmatrix}
     \a \\ c \\ d \\ uJ_\ch
  \end{bmatrix}_{x+L}
  \hspace*{-2mm}
  = 
  \begin{bmatrix}
   1 & 0 & 0 & \frac{ue}{\sigma S} L
   \\
   0 & e^{-L/\lambda} & 0 & 0
   \\
   0 & 0 & e^{L/\lambda} & 0
   \\
   0 & 0 & 0 & 1
  \end{bmatrix}
  \cdot
  \begin{bmatrix}
    \a \\ c \\ d \\ uJ_\ch
  \end{bmatrix}_{x}
  \notag
  \\[1mm]
  \displaystyle
  \vek{F}_{x+L}=[\mat{P}]^{-1}\cdot\vek{F}_x.
  \label{eq:defP}
\end{gather}%
Hence, the relation between
$\vek{H}=[\mu_\up,\,\mu_\dw,\,uJ_\up,\,uJ_\dw]$ at both ends of the
layer is expressed as
\begin{equation}
\label{eq:defk}
  \displaystyle
  \vek{H}_{x}=\mat{D} \mat{P} \mat{D}^{-1} \vek{H}_{x+L}
  =\mat{K} \vek{H}_{x+L}.
\end{equation}%

\subsection{Interface resistivity and shunting interface resistance}
\label{s:int}

In the previous Section, we have expressed response of one layer in
homogeneous material. Here, we describe the interfacial
properties including (i) interface resistance \cite{val93} $AR_{\up/\dw}=2
AR^\star (1\mp\gamma)$ and (ii) interfacial spin-flip scattering,
described by a~shunting resistance $AR_s$, ``shortcutting'' up and
down channels at the interface. The response of
interface can be expressed by $\mat{K}_\mathrm{int}$ matrix
\begin{equation}
\label{eq:Kint0}
  \vek{H}_{x-\mathrm{d}x}=\mat{K}_\int \vek{H}_{x+\mathrm{d}x},
\end{equation}
where $x\mp\mathrm{d}x$ denotes for $\vek{H}$-vector just below/above
interface and  $\mat{K}_\int$ writes
\begin{equation}
\label{eq:Kint}
  \mat{K}_\int=\begin{bmatrix}
    1 & 0 & -\frac{AR_\up}{uS} & 0 \\
    0 & 1 & 0 & -\frac{AR_\dw}{uS} \\
    -\frac{uS}{AR_s} & \frac{uS}{AR_s} & 1 & 0\\
    \frac{uS}{AR_s} & -\frac{uS}{AR_s} & 0 & 1
  \end{bmatrix}.
\end{equation}%
Notice that alternative interfacial spin-flip scattering by
$\delta_I$-parameter was introduced in \cite{par00,mar01} describing
spin relaxation at the interfaces by a~thin interfacial layer of
spin-flip-length $\lambda_I=t_I/\delta_I$ where $t_I$ is interfacial
layer thickness. From comparison of
$\mat{K}$-matrices from Eqs.(\ref{eq:defk}) and (\ref{eq:Kint}),
$AR_s=4 t_I/(\sigma_I \delta_I^2) \mathrm{sinh}\,\delta_I$ where
$t_I$ and $\sigma_I$ are interfacial layer thickness and conductivity,
respectively. When interfacial spin-flip scattering does not exist,
then $R_s=\infty$.

\subsection{Simple multilayer structure}
\label{s:multilayer}

Although the formalism developed here is mainly to calculate
$\mu_\updw$, $j_\updw$ inside an electrical circuit, we shall first
show how presented $4\times4$ matrix algebra can be used to calculate
electrical response of a~multilayer structure [Fig.~\ref{f:sdre}(c)].

The description for a single layer by matrix $\mat{K}$ in
Eq.~(\ref{eq:defk}) gives relationship between $\mu_\updw$ and
$J_\updw$ at both ends of the layer.  Hereafter, instead of continuity
of spin-polarized current density $j_\updw$ over interface, we consider
continuity of spin-polarized current $J_\updw$, to take into account
variable cross-sectional area $S$ of layers.  The reasons and
validity are discussed later in Section~\ref{s:cr1dm}.
Obviously, when all layers has the same $S$, our formalism provides
the same results as original VF formalism \cite{val93}.

The boundary conditions at interfaces are continuity of $\mu_\updw$
and $J_\updw$, i.e. continuity of $\vek{H}$-vector.  Consequently, the
response of whole multilayer structure can be written as \cite{yeh80,vis86}
\begin{equation}
\label{eq:defM1}
  \vek{F}^{(0)}=\mat{M}^{(M+1)}
  \cdot \vek{F}^{(M+1)},
\end{equation}%
where
\begin{multline}
\label{eq:defM}
  \mat{M}^{(M+1)}=
  [\mat{D}^{(0)}]^{-1} \, 
  \mat{K}^{(0)}_\int\, \mat{K}^{(1)}\, \mat{K}^{(1)}_\int 
  \ldots \\
  \mat{K}_\int^{(M-1)}
  \mat{K}^{(M)}\, \mat{K}^{(M)}_\int\,
 \mat{D}^{(M+1)} 
\end{multline}
where upper index in parenthesis denotes for interface or layer
number, $M$ the number of layers. 

Because materials $(0)$ and $(M+1)$ are semi-infinite, $\mu_\updw$
inside them must not exponentially increase. Hence, some
exponential terms in Eq.~(\ref{eq:defmu}) must vanish, namely $c^{(0)}\equiv0$,
$d^{(M+1)}\equiv0$.  Hence, the vectors $\vek{F}^{(0)}$ and
$\vek{F}^{(M+1)}$ are limited to form
\begin{equation}
\label{eq:endF}
\vek{F}^{(0)}=
\begin{bmatrix}
  \a^{(0)} \\ 0 \\ d^{(0)} \\ uJ_\ch
\end{bmatrix}
\qquad
\vek{F}^{(M+1)}=
\begin{bmatrix}
  \a^{(M+1)} \\ c^{(M+1)}\\ 0 \\ uJ_\ch
\end{bmatrix}.
\end{equation}%

Now, Eq.(\ref{eq:defM1}) can be solved. 
Substituting Eq.~(\ref{eq:endF}) to Eq.~(\ref{eq:defM1}), we
found all unknowns in $\vek{F}^{(M+1)}$
\begin{equation}
\label{eq:ac}
\begin{bmatrix}
  \a^{(M+1)} \\ c^{(M+1)}
\end{bmatrix}
=
\left(
\begin{bmatrix}
  M_{11} & M_{12} \\
  M_{21} & M_{22}
\end{bmatrix}
\right)^{-1}
\cdot
\begin{bmatrix}
  -uM_{14}J_\ch + \a^{(0)} \\ -uM_{24}J_\ch
\end{bmatrix},
\end{equation}%
where $M_{ij}$ are elements of $\mat{M}$-matrix.  The value of
$\a^{(0)}$ can be arbitrary value, as it only adds a constant to the
profiles of $\mu_\updw$.  Other simplification of Eq.~(\ref{eq:ac})
follow from $M_{11}\equiv1$ and $M_{21}\equiv0$, in consequence of (i)
for $J_\ch=0$ the terms $\mu_\updw$, $\a$ must be constant and equal to each
other and (ii) for $J_\ch=0$, $c^{(M+1)}=0$.  Then, solution of
Eq.~(\ref{eq:ac}) can be written as
\begin{equation}
\label{eq:ac1}
\begin{array}{rl}
\displaystyle
\a^{(M+1)}-\a^{(0)}&
\displaystyle
=\left(-M_{14} + \frac{M_{12}M_{24}}{M_{22}}\right)uJ_\ch\\
\displaystyle
c^{(M+1)}&
\displaystyle
=-\frac{M_{24}}{M_{22}}uJ_\ch.
\end{array}
\end{equation}%
Now, $\a^{(M+1)}$ and $c^{(M+1)}$ are known and therefore the vector
$\vek{F}^{(M+1)}$ can be reconstructed from Eq.~(\ref{eq:endF}).
Consequently, the profiles of $\mu_\updw$, $\a$, $J_\updw$ etc.\ in
the entire structure can be determined by recursive applying
step-by-step matrix multiplications in Eq.~(\ref{eq:defM})

Finally, resistivity of the multilayer structure (between
first and last interface) is $R=(\a^{(M+1)}-\a^{(0)})/J_\ch$.

\section{Electric circuit}
\label{s:net}

As demonstrated in the previous Section, the $\mat{M}$-matrix [defined
by Eq.~(\ref{eq:defM1})] can characterize the entire multilayer
structure where
the same charge current $J_\ch$ flows across all layers. In this Section,
we extend the previous formalism to an electrical circuit (network) of
SDREs, mutually connected at nodes. In general, the SDRE is composed
of any sequence of materials (layers) and/or interfaces, as depicted
on Fig.~\ref{f:sdre}. There are three types of SDRE, depending whether
the length of SDRE is finite or infinite:
\begin{itemize}
\item close-end SDRE [Fig.~\ref{f:sdre}(a)] has finite length and
  hence both
  ends of this SDRE are attached to nodes. Because the boundary
  condition at nodes are described by $\vek{H}$-vectors, whole SDRE is
  described by $\mat{K}^{[b]}$-matrix relating $\vek{H}$-vectors at
  both ends of SDRE:
  \begin{equation}
  \label{eq:closeend}
  \vek{H}^{[b](0)}=\mat{K}^{[b]}\vek{H}^{[b](M_b)},
  \end{equation}
  where, for later purpose, $\mat{K}^{[b]}$ can be rewritten into four
  $2\times2$ submatrices
  \begin{equation}
  \label{eq:closeend1}
    \begin{bmatrix}
    \vek{\mu}\\ u\vek{J}
  \end{bmatrix}^{[b](0)}=
  \begin{bmatrix}
    \mat{K}_{\mu\mu} & \mat{K}_{\mu J} \\
    \mat{K}_{J\mu} & \mat{K}_{JJ}
  \end{bmatrix}^{[b]}
  \begin{bmatrix}
    \vek{\mu}\\ u\vek{J}
  \end{bmatrix}^{[b](M_b)},
  \end{equation}
  where $b$ denotes a SDRE number in the circuit and $M_b$ is the
  number of layers in $b$-th SDRE,
  $\vek{\mu}^{[b](M_b)}\equiv
  [\mu^{[b](M_b)}_{\up},\mu^{[b](M_b)}_{\dw}]^\mathrm{T}$
  and
  $\vek{J}^{[b](M_b)}\equiv[J^{[b](M_b)}_{\up},J_\dw^{[b](M_b)}]^\mathrm{T}$,
  $^\mathrm{T}$ denoting vector transposition.
  Hereafter, indices in square bracket denotes for SDRE number,
  whereas indices in ordinary parentheses denotes for number of layer
  inside SDRE.  Positive current direction is from layer (1) to layer
  $(M_b)$.  Analogous to Eq.~(\ref{eq:defM}), $\mat{K}^{[b]}$ consists
  of layer and interface contributions
  \begin{equation}
    \label{eq:defk0}
    \mat{K}^{[b]}= \mat{K}^{[b](1)}\, \mat{K}^{[b](1)}_\int 
    \ldots \mat{K}_\int^{[b](M_b-1)} \mat{K}^{[b](M_b)}.
  \end{equation}
  When SDRE contains only one layer (i.e.\ it consists of single
  metal), then [Eq.~(\ref{eq:defk})] $\mat{K}^{[b]}=\mat{K}^{[b](1)}\equiv
  \mat{D}^{[b]}\mat{P}^{[b]}[\mat{D}^{[b]}]^{-1}$.
\item open-end SDRE [Fig.~\ref{f:sdre}(b)] has one end ``finite'' and
  connected to the node.  The other end is infinitely long and at its
  end the charge current $J_\ch^{[b](0)}$, flowing into the $b$-th open-end
  SDRE, is applied.  Because boundary conditions on node are described
  by $\vek{H}$-vector and boundary conditions of the propagation into
  infinity by $\vek{F}$-vector, the $b$-th open-end SDRE is described
  by $\mat{Z}^{[b]}$-matrix as
  \begin{equation}
  \label{eq:openend}
  \vek{F}^{[b](0)}=\mat{Z}^{[b]}\vek{H}^{[b](M_b)},
  \end{equation}
  where we have used convention that direction of SDRE and positive
  current direction goes from infinite end of SDRE toward its finite
  end [Fig.~\ref{f:sdre}(b)]. Analogous to Eqs.~(\ref{eq:defM}) and
  (\ref{eq:defk0})
  \begin{multline}
    \label{eq:defz0}
    \mat{Z}^{[b]}=[\mat{D}^{[b](0)}]^{-1} 
    \mat{K}_\int^{[b](0)}
    \mat{K}^{[b](1)}\,
    \mat{K}^{[b](1)}_\int  
    \ldots 
    \\
    \mat{K}_\int^{[b](M_b-1)} \mat{K}^{[b](M_b)}.
  \end{multline}  
  When the SDRE contains no layer (i.e. it contains only single
  material continuous to infinity), then
  $\mat{Z}^{[b]}=[\mat{D}^{[b]}]^{-1}$.
\item %
  multilayer structure [Fig.~\ref{f:sdre}(c)] described in
  Sect.~\ref{s:multilayer} may be understood as a~special type of
  SDRE, having both ends open-ended.
\end{itemize}

In general, the electric circuit is assumed to have $N$ nodes, $C$
close-end SDRE and $E$ open-end SDRE.

The boundary conditions valid for each node are determined by generalized
Kirchoff's laws:
\begin{gather}
\label{eq:kirmu}
\mu_{n,\up/\dw}^{[b_n]}=\mathrm{const}_{n,\up/\dw}\equiv \mu_{n,\updw},
\\
\label{eq:kirJ}
\sum_{b_n=1}^{B_n} J_{n,\up/\dw}^{[b_n]}=0,
\end{gather}%
i.e. the values of $\mu_\updw$ has to be identical for all ends of
SDRE connected to each node and sum of polarized currents $J_\updw$
entering each node has to be zero.  Subscript $n$ denotes for node
number, $n=1\ldots N$ and $b_n=1,\ldots B_n$ denotes the SDREs
connected to the $n$-th node. 

Hereafter, we use two $\mu$-notation (i) $\mu_{n,\updw}$ relating
$\mu_\updw$ at the $n$-th node and (ii) $\mu^{[b](0)}_\updw$,
$\mu^{[b](M_b)}_\updw$ denoting $\mu_\updw$ at start, end of the
$b$-th SDRE, respectively. The relation between these two
$\mu$-notations is given by connections between nodes and SDREs, i.e.
by the topology of the electric circuit.

Following the previous discussion, the problem is how to treat large
number of linear equations
(\ref{eq:closeend})(\ref{eq:openend})(\ref{eq:kirmu})(\ref{eq:kirJ})
giving relations between $\mu_\updw$ and $J_\updw$ on the nodes and
between ends of SDREs. We do it by solving one large matrix
expression,
\begin{equation}
\label{eq:largeQ}
  \mmat{Q}\cdot\vvek{H}=\vvek{F},
\end{equation}%
to which we apply all the previously mentioned rules.  The
$\vvek{H}$-vector contains $\mu_{n,\updw}$ for all nodes and also
$uJ_\updw^{[b](M_b)}$, $b=1\ldots C+E$, at the \textit{end} (i.e.\ at
the $M_b$-side) of each SDRE. Hence, $\vvek{H}$-vector has $2(C+E+N)$
elements.

The role of the $\mmat{Q}$-matrix consists of three parts
[Eq.(\ref{eq:Qex})]: 
\begin{itemize}
\item[(1)] to relate $\mu_\updw$, $J_\updw$ at the ``finite'' end of
  the open-end SDREs and between boundary conditions of propagation
  towards infinity. This is provided by $2E$ linear equation, and
  hence this part occupies $2E$ rows in $\mmat{Q}$-matrix.
\item[(2)] to relate $\mu_\updw$ between start and end of the
  close-end SDREs ($2C$ linear equations).
\item[(3)] to realize current conservation at each node ($2N$ linear
  equations).
\end{itemize}
These contributions are studied in detail in following. 
In total,
$\mmat{Q}$-matrix has $2(C+E+N)$ rows, the same as length of
$\vek{H}$-vector. So $\mmat{Q}$ is a~square matrix.

\figII

An example of a circuit of SDREs is depicted in Figure~\ref{f:netsch}.
The circuit consist of $N=3$ nodes, connected by SDRE.  The circuit
consists of $C=3$ close-end SDRE and $E=2$ open-end SDREs.  Then, the
resulting equation $\mmat{Q}\vvek{H}=\vvek{F}$ looks like:
\begin{widetext}
\begin{equation}
\label{eq:Qex}
\left[
\begin{array}{ccc|ccccc}
\tilde{\mat{Z}}^{[1]}_\mu & \mat{0} & \mat{0} & \tilde{\mat{Z}}^{[1]}_J & 
  \mat{0} & \mat{0} & \mat{0} & \mat{0}
  \\
\mat{0} & \mat{0} & \tilde{\mat{Z}}'^{[5]}_\mu 
 & \mat{0} & \mat{0} & \mat{0} & \mat{0} & \tilde{\mat{Z}}'^{[5]}_J
\\
\hline
-\mat{1} & \mat{K}^{[2]}_{\mu,\mu} & \mat{0} & \mat{0} &
  \mat{K}_{\mu J}^{[2]} & \mat{0} & \mat{0} & \mat{0}
\\
\mat{0} & -\mat{1} & \mat{K}^{[3]}_{\mu,\mu} & \mat{0} &\mat{0} & 
  \mat{K}_{\mu J}^{[3]} & \mat{0} & \mat{0}
\\
-\mat{1} & \mat{0} & \mat{K}^{[4]}_{\mu,\mu} & \mat{0} &\mat{0} & 
  \mat{0} & \mat{K}_{\mu J}^{[4]} & \mat{0} 
\\
\hline
\mat{0} & \mat{K}^{[2]}_{J\mu} & \mat{K}^{[4]}_{J\mu} & -\mat{1} 
  & \mat{K}^{[2]}_{JJ} & \mat{0} & \mat{K}^{[4]}_{JJ} & \mat{0}
\\
\mat{0} & \mat{0} & \mat{K}^{[3]}_{J\mu} & \mat{0} & -\mat{1} &
  \mat{K}^{[3]}_{JJ} & \mat{0} & \mat{0}
\\
\mat{0} & \mat{0} & \mat{0} & \mat{0} & \mat{0} & 
  -\mat{1} & -\mat{1} & -\mat{1} 
\end{array}
\right]
\begin{bmatrix}
\vek{\mu}_1
\\
\vek{\mu}_2
\\
\vek{\mu}_3
\\
\hline
u\vek{J}^{[1](M_1)}
\\
u\vek{J}^{[2](M_2)}
\\
u\vek{J}^{[3](M_3)}
\\
u\vek{J}^{[4](M_4)}
\\
u\vek{J}^{[5](M_5)}
\end{bmatrix}
=
\begin{bmatrix}
\tilde{\vek{F}}^{[1](0)} \\
\tilde{\vek{F}}'^{[5](0)} \\
\hline
\vek{0}\\ \vek{0}\\ \vek{0} \\
\hline
\vek{0} \\\vek{0} \\ \vek{0}
\end{bmatrix}
\end{equation}%
\end{widetext}
where $\vek{\mu}_n\equiv[\mu_{n,\up},\mu_{n,\dw}]^\mathrm{T}$ and
$\mat{1}$, $\mat{0}$ denote $2\times2$ unitary,
zero matrix, respectively. Each row which is shown in
$\mmat{Q}$-matrix in Eq.~(\ref{eq:Qex}) represents two rows (for up
and down channel), so hereafter we call it ``double-row''.

(ad 1) The role of the first part in the $\mmat{Q}$-matrix relates
both ends of open-end SDRE, described by Eq.~(\ref{eq:openend}),
$\vek{F}^{[b](0)}=\mat{Z}^{[b]}\vek{H}^{[b](M_b)}$.  However, we only
know 2 variables out of 4 in vector $\vek{F}^{[b](0)}=
[\a^{[b](0)},
c^{[b](0)}, d^{[b](0)}, uJ_\ch^{[b](0)} ]^\mathrm{T}$.
We know charge current
$J_\ch^{[b](0)}$ entering $b$-th SDRE and that $c^{[b](0)}=0$, because
$\mu_\updw$ must not exponentially increase towards infinity
[Eq.~(\ref{eq:endF})]. Hence, taking from $\mat{Z}^{[b]}$-matrix
[Eq.(\ref{eq:openend})] only rows corresponding with known value in
$\vek{F}^{[b](0)}$, we get
\begin{multline}
  \label{eq:deftildeZ}
\hspace*{-6mm}
  \tilde{\vek{F}}^{[b](0)}\!\!\equiv\!\!
  \begin{bmatrix}
    0 \\ uJ_\ch^{[b](0)}  
  \end{bmatrix}
  \!\!=\!\!
  \begin{bmatrix}
    Z_{21}\! &\!Z_{22}\! &Z_{23}\! &\!Z_{24} \\
    Z_{41}\! &\!Z_{42}\! &Z_{43}\! &\!Z_{44} 
  \end{bmatrix}^{[b]}
  \!
  \begin{bmatrix}
    \mu_\up \\ \mu_\dw \\ uJ_\up \\ uJ_\dw
  \end{bmatrix}^{[b](M_b)}
  \\ \equiv \begin{bmatrix}
  \tilde{\mat{Z}}^{[b]}_{\mu} &   \tilde{\mat{Z}}^{[b]}_{J}
  \end{bmatrix}
  \begin{bmatrix}
    \vek{\mu}^{[b](M_b)} \\ u\vek{J}^{[b](M_b)}
  \end{bmatrix}.
\end{multline}%
This equation is substituted into first part of in
$\mmat{Q}$-matrix 
[Eq.~(\ref{eq:Qex})] in the form: $\tilde{\mat{Z}}^{[b]}_\mu
\vek{\mu}^{[b](M_b)} + \tilde{\mat{Z}}^{[b]}_J u \vek{J}^{[b](M_b)}
=\tilde{\vek{F}}^{[b](0)}$.

In the case of the last open-end SDRE connected to the structure (in
our example SDRE number $b'=5$), the situation is slightly different.
In the vector $\vek{F}^{[b'](0)}= [ \a^{[b'](0)}, c^{[b'](0)},
d^{[b'](0)}, uJ_\ch^{[b'](0)} ]^\mathrm{T}$ (a) we have to set a value of
$\a^{[b'](0)}$, which sets an absolute value of all $\mu_{\updw}$ and $\a$
inside circuit to an arbitrary value and (b) getting a charge current
$J_\ch^{[b'](0)}$ would be redundant as a~sum of all charge currents
entering circuit has to be zero. Hence, the last double-row in the
first part of the $\mmat{Q}$-matrix (second double-row in
Eq.~(\ref{eq:Qex})) looks like
\begin{multline}
  \label{eq:defprimeZ}
  \hspace*{-6mm}
  \tilde{\vek{F}}'^{[b'](0)}
  \!\!\equiv\!\!
  \begin{bmatrix}
    \a^{[b'](0)}\\ 0   
  \end{bmatrix}
  \!=\!
  \begin{bmatrix}
    Z_{11}\! &\!Z_{12}\! &\!Z_{13}\! &\!Z_{14} \\
    Z_{21}\! &\!Z_{22}\! &\!Z_{23}\! &\!Z_{24} 
  \end{bmatrix}^{[b']}
  \!
  \begin{bmatrix}
    \mu_\up \\ \mu_\dw \\ uJ_\up \\ uJ_\dw
  \end{bmatrix}^{[b'](M_b)}
  \\ \equiv \begin{bmatrix}
  \tilde{\mat{Z}}'^{[b']}_{\mu} &   \tilde{\mat{Z}}'^{[b']}_{J}
  \end{bmatrix}
  \begin{bmatrix}
    \vek{\mu}^{[b'](M_b)} \\ u\vek{J}^{[b'](M_b)}
  \end{bmatrix}
\end{multline}%
and is substituted to $\mmat{Q}$-matrix analogous as
Eq.~(\ref{eq:deftildeZ}).

(ad 2)
The second part in the $\mmat{Q}$-matrix gives relation between
$\mu_{\updw}^{[b](0)}$ at the start of the $b$-th close-end SDRE and
$\mu_{\updw}^{[b](M_b)}$, $J_{\updw}^{[b](M_b)}$ at the end of $b$-th
SDRE. This relation is given by the first double row taken from
Eq.~(\ref{eq:closeend1}). It is substituted into $\mmat{Q}$  in the form
$0=-\vek{\mu}^{[b](0)}+\mat{K}^{[b]}_{\mu\mu}\vek{\mu}^{[b](M_b)}
+u\mat{K}^{[b]}_{\mu J}\vek{J}^{[b](M_b)}$.
  
(ad 3) The last part in $\mmat{Q}$-matrix describes the current
conservation at each node, described by Kirchoff's law
Eq.~(\ref{eq:kirJ}).
As specified, $\vvek{H}$-vector contains only
values of current at the end of each SDRE, $J_\updw^{[b](M_b)}$.
Hence, if the $b$-th SDRE starts at the $n$-th node (as No.~2 and~4 SDREs
at the node~1), then the current is expressed by the second double-row
in Eq.~(\ref{eq:closeend1}), from $\mu_\updw^{[b](M_b)}$,
$J_\updw^{[b](M_b)}$ as
$u\vek{J}^{[b](0)} =
\mat{K}^{[b]}_{J\mu} \vek{\mu}^{[b](M_b)}
+u\mat{K}_{JJ}^{[b]}\vek{J}^{[b](M_b)}$.
For example, in case of node $n=1$, the current conservation is 
$-J_\updw^{[1](M_1)}+J_\updw^{[2](0)}+J_\updw^{[4](0)}=0$, which is
substituted to 
$\mmat{Q}$-matrix to the 6-th double-row as 
$-\vek{J}^{[1](M_1)}
+
\mat{K}^{[2]}_{J\mu} \vek{\mu}^{[2](M_2)}
+u\mat{K}_{JJ}^{[2]}\vek{J}^{[2](M_2)}
+
\mat{K}^{[4]}_{J\mu} \vek{\mu}^{[4](M_4)}
+u\mat{K}_{JJ}\vek{J}^{[4](M_4)}=\mat{0}$.

Although the construction of the $\mmat{Q}$-matrix as presented here may
be tedious, it is rather direct to establish its construction 
numerically. When the equation $\mmat{Q}\vvek{H}=\vvek{F}$ is solved,
values of $\mu_{\updw}$, $J_\updw$ for each SDRE
are directly written in $\vvek{H}$-vector; their profiles can be
determined by step-by step applying multiplication in
Eqs.(\ref{eq:closeend})(\ref{eq:defk0}) and
(\ref{eq:openend})(\ref{eq:defz0}).

\subsection{Construction of 3D electric circuit}
\label{s:grid}

In the previous Section we have derived the formalism to calculate
$J_\updw$ and $\mu_\updw$ in arbitrary electric circuit. In this
Section, we explain, how to describe the whole nano-structure as a
circuit of SDREs, as sketched in Fig.~\ref{f:grid}.

\figIII

Each part of the nanostructure is divided into 3D rectangular grid,
the circuit nodes positions being $x_i$, $y_j$, $z_k$.  Then, a given
SDRE, for example in the $x$-direction, has length
$L_i=(x_{i-1}+x_i)/2$ and cross-sectional area $S_{jk}=y_j z_k$. Due
to this treatment, the grid does not need to be equally spaced, but
just rectangular. 

At the interface between two different materials, e.g.\ A and B (see
Fig.~\ref{f:grid}), SDRE is described by $\mat{K}_{A\rightarrow B}$
matrix consisting of three contributions $\mat{K}_{A\rightarrow
  B}=\mat{K}_{A\rightarrow\int} \mat{K}_\int \mat{K}_{\int\rightarrow
  B}$ [Eq.~(\ref{eq:defk0})]. The $\mat{K}_{A\rightarrow\int}$ is
contribution from grid point (node) inside A material to the interface
with (SDRE is now along the z-direction) $L_k=z_\int-z_{k-1}$ and
$S_{ij}=x_i y_j$. $\mat{K}_\int$ is the interface resistivity matrix
given by Eq.(\ref{eq:Kint}).  $\mat{K}_{\int\rightarrow B}$ is
contribution from the interface to the node in B material having
$L=z_{k}-z_\int$ and $S_{ij}=x_i y_j$.
 
The electrical circuit network as described above describes correctly
charge current. However, inside 3D (2D) SDRE circuit the
volume of metal in all resistors is three (two) times larger than in
reality.  In such a case, the spin-polarized current would have larger
dumping, as it would diffuse into larger volume. To correct this, it
is necessary to calculate 1D propagation of $\mu_{\updw}$ in each
SDRE by slightly modified Eq.~(\ref{eq:diffmu})
\begin{equation}
  \label{eq:diffmudim}
  f \frac{\partial^2 (\mu_\up-\mu_\dw)}{\partial x^2}
    =\frac{\mu_\up-\mu_\dw}{\lambda^2},
\end{equation}
where $f$ is a~dimension of the SDRE circuit. In other words, when
a~given nanostructure is described by 3D (2D)
electrical circuit, the $\lambda$ should be increased by a
factor $\sqrt{3}$ ($\sqrt{2}$). This $\lambda$-normalization
should not be applied for open-end SDREs, as their contribution is 
correctly described by 1D propagation.
 
\figIV
\figV

The advantage of this $\lambda$-normalization is shown in
Figure~\ref{f:2d3d}, where $j_\sp$ through (Cu/Co)$^2$ structure
(described and studied in detail in the next Section) is compared between
1D model (full line), and our 3D model without
$\lambda$-normalization ($\circ$) and with $\lambda$-normalization
($\times$). We can see that with (without) $\lambda$-normalization,
the agreement between 1D and 3D model is about 1\% (30\%). The same is
valid for spin-accumulation $\Delta\mu=\mu_\up-\mu_\dw$, where
calculation without $\lambda$-normalization leads to 30\% smaller
$\Delta\mu$.

\figVI

\subsection{Surface scattering}
\label{s:surf}

In this Section, we describe how to incorporate surface scattering 
to the presented formalism.

Surface scattering can be described by a shunting resistance $R_s$
shortcutting up and down channels for a nodes situated just close to the
wire surface. $R_s$ has value $R_{s,n}=AR_s/S_n$, $AR_s$ being surface
scattering resistivity and $S_n$ being surface area corresponding to
the $n$-th node. When surface scattering is not presented, then
obviously $R_{s,n}=\inf$. 
Surface scattering can be described by modification of Kirchoff law
[Eq.~(\ref{eq:kirJ})]
\begin{equation}
\label{eq:kirJ1}
\sum_{k_n} J_{n,\up/\dw}^{[k_n]}\mp \frac{\mu_{n,\up}-\mu_{n,\dw}}{R_s}=0.
\end{equation}

To incorporate this modification into $\mmat{Q}$-matrix for the $n$-th node,
we add $\mat{G}_s$ matrix
\begin{equation}
  \label{eq:rs}
  \mat{G}_s=\frac{1}{R_s}
  \left[
  \begin{array}{rr}
    -1 &1 \\ 1 & -1 
  \end{array}
  \right]
\end{equation}
on the $n$-th double-column and $n$-th double-row in the last part of
$\mmat{Q}$-matrix which describes current conservation in each node. In our
example given by Eq.~(\ref{eq:Qex}) and Figure~\ref{f:netsch}, to add
$R_s$ to $n=2$ node, we place $\mat{G}_s$ to the 7-th double-row and
2-nd double-column into $\mmat{Q}$-matrix.

\section{Application to (C\lowercase{u}/C\lowercase{o}) pillar structure}
\label{s:cuco}

\tabI

In this Section we use the above developed formalism on (Cu/Co)$^2$
and (Cu/Co)$^3$ pillar structures.  We will show how spin-polarized
current density $j_\sp=j_\up-j_\dw$ and electrochemical potential
$\mu_\updw$ vary differently between when whole structure is a pillar
or only a~part of the structure is a~pillar attached to an infinitely
large continuous layer. In the literature, pillars terminated with
infinitely large continuous layers are commonly used
\cite{alb00,kat00,alb02,gro03,sun03}.

We have studied structures consisting of 2 and 3 Co layers, called
(Cu/Co)$^2$, and (Cu/Co)$^3$ with dimensions in nm
Cu1/Co1(40)/Cu2(6)/Co2(2)/Cu3 and
Cu1/Co1(40)/Cu2(6)/Co2(2)/Cu3(6)/Co3(20)/Cu4, respectively.  The
square-shaped pillar 100\,nm in size begins with the Co1/Cu2
interface. The considered structure types are defined in
Fig.~\ref{f:types} as (a) cross-sectional area S constant, (b) column,
(c) constriction and (d) constriction with infinite Co1 layer.  In the
following discussion, ``constriction'' corresponds to the case (c).
The ``infinite'' homogeneous layers were approximated as a~square
pillar of 800\,nm in size.  The magnetization of Co1 layer is always
fixed as ``up'' ($\up$), whereas magnetic orientations of other Co
layers are varied.  Note that in our diffusive transport calculations, the
magnetization orientation with respect to the structure (e.g. in-plane
or out-of-plane) do not play any role, but only mutual magnetization
orientation (parallel or antiparallel) does play an important role.
The charge current passing through structure is assumed to be $J_\ch=1$\,mA,
equivalent to the averaged charge current density in the pillar
$j_\ch=100\times10^{9}$\,A/m$^2$.  The electrical properties of
materials are of room temperature 
\cite{jed03,upa98,pir98}: electric conductivity
$\sigma_\cu=4.81\times10^6$\,$\Omega^{-1}$m$^{-1}$,
$\sigma_\co=4.2\times10^6$\,$\Omega^{-1}$m$^{-1}$, spin-flip-lengths
$\lambda_\cu=350$\,nm, $\lambda_\co=60$\,nm and Co spin bulk assymetry
$\gamma=0.35$. We assume no interface resistance and no interface and
surface scattering. The SDRE grid size is 10\,nm.

\subsection{Current density in the structure}
\label{s:cr-through}

Figure~\ref{f:crsp} shows the profile of $j_\sp$ along center axis of
the structures: (a) (Cu/Co)$^2$ (b) (Cu/Co)$^3$ with parallel Co1 and
Co3 layers and (c) (Cu/Co)$^3$ with antiparallel Co1 and Co3 layers
for S constant, column and constriction type structures.  In all
cases, $j_\sp$ for parallel Co1 and Co1 layer is larger than for
antiparallel configuration.  Furthermore, $j_\sp$ is enhanced at the
position of free Co2 layer for the column and constriction types
compared to the S~constant type structure.  For example, in
the case of the constriction type structure with (Cu/Co)$^2$ $\up\up$
configuration, $j_\sp$ is enhanced by a~factor of 1.75 and in the case
of (Cu/Co)$^3$ $\up\up\up$ by 1.5. When only Cu buffer layer is
infinite in size (column-type structure), $j_\sp$ at the position of
Co2 layer is enhanced too, but not so strongly.
Figure~\ref{f:crsp}(c) shows that for (Cu/Co)$^3$ with antiparallel
Co1 and Co3 layers, $j_\sp$ is significantly reduced for all types of
structures.

The origin of the $j_\sp$ enhancement is following: the pillar is
attached to infinitely large Cu layer which provides large volume for
spin-current to be scattered and so acting as a~strong
spin-flip-scatterer.  Hence, infinitely large Cu layer works as
a~small shortcutting resistance between up and down channels.
Consequently, shortcutting of up and down channels leads to an
increase in $j_\sp$. The increase of $j_\sp$ is related with increase
of spin-polarization efficiency $p=j_\sp(J_\ch/S_\mathrm{pillar})$, as
charge current flowing though pillar $J_\ch$ is fixed in all our
calculations. Consequently, increase of $p$ leads to decrease of
critical switching current $J_{s,\ch}$ which is necessary to reverse
magnetization direction of free layer.

As the constriction type structure is in the most common use for Co/Cu
pillar structures, the $j_\sp$ enhancement (i.e.\ $J_{s,\ch}$ reduction) has
been already widely used \cite{sun03,oez03} without being noticed.
Similar effect can be realized by inserting a~layer with small
characteristic spin-flip resistance $AR_\lambda=\lambda/\sigma$, such
as Pt, Ag, Au, Ru above the last Co layer or bellow the first Co
layer.  Such cover layers has been used \cite{eml04,jia04} since the
first pioneering experiment \cite{kat00}, but their contributions to
the $j_\sp$ enhancement have been observed recently
\cite{jia04,ura03a, ura03b}.

\figVII
\figVIII
\figIX

Figure~\ref{f:crch} shows profiles of charge current
$j_\ch=j_\up+j_\dw$ along the center axis of the (Cu/Co)$^3$ structure
for S constant, column and constriction type structures. Obviously,
for S constant structure, the $j_\ch$ is constant. In the case of
infinite Cu termination, $j_\sp$ decreases approximately exponentially
over the characteristic length of 50\,nm. The same decay of $j_\sp$ is
presented in Fig.~\ref{f:crsp} for infinitely large Cu layers.

In Table~\ref{t:tab}, we summarize averadged values of $j_\sp$ in the
position of Co2 layer in all the types of studied structures. These
$j_\sp$ values may differ from those presented in Fig.~\ref{f:crsp}
due to lateral inhomogenity of $j_\sp$ inside pillar, discussed in
next Section. The largest averaged $j_\sp$ is obtained for
(Cu/Co)$^3$ $\up\up\up$ constriction structuture
($31.0\times10^{9}$\,A/m$^2$) and (Cu/Co)$^2$ $\up\up$ constriction
structure ($29.4\times10^9$\,A/m$^2$), providing $j_\sp$ enhancement
by a~factor~2 with respect to (Co/Cu)$^2$ S-constant structure
(14.85$\times 10^9$\,A/m$^2$). This tendency is already explained in
above paragraph.

\subsection{Current inhomogenity inside pillar}
\label{s:crin}

\figX

Figure~\ref{f:pl2} presents a map of spin-polarized current density
$j_\sp$ inside (Cu/Co)$^2$ constriction type structure for (a)(b)
parallel and (c) antiparallel magnetization configuration. The case (a)
is a structure without infinitely large Co1 layer whereas (b)(c) with
it.  A~map of $j_\ch$ is not presented here as it looks
similar to $j_\sp$ with $\up\up$. The $j_\sp$ inside Co2 layers is more
homogenous and flows well perpendicular to the interfaces
although in the surrounding Cu layers the $j_\sp$ can have rather
large inclination and inhomogenity. The $j_\sp$ tends to be more
homogeneous when passing Co layer, causing that the $j_\sp$ in the
adjacent Cu layers can have rather large in-plane components.  This
is remarkable in the case (c). Very similar tendency is 
found for $j_\ch$.

The above mentioned characteristics of $j_\sp$ are consequence of
larger spin-flip resistance $AR_\lambda=\lambda/\sigma$ of Co with
respect to Cu, $AR_{\lambda,\mathrm{Co}}$=14.3\,f$\Omega$m$^2$,
$AR_{\lambda,\mathrm{Cu}}$=7.3\,f$\Omega$m$^2$. It means that
spin-flip is more likely to occur inside Cu than inside Co. In other
words, Co is ``harder'' material than Cu for $j_\sp$ to penetrate into
it. The above mentioned characteristics of $j_\ch$ are simply
consequence of $\sigma_\co\ll\sigma_\cu$.

Figure~\ref{f:prftra} shows cross-sectional (in $x$-direction) profiles of
(a) $j_\sp$ and (b) $j_\ch$ through pillar in the middle of the Co2
layer for (Cu/Co)$^2$ structure. Both $j_\sp$ and $j_\ch$ are
inhomogenous having minima at the structure center.  This is due to
inhomogenous current injection into the pillar from infinitely large
layers.  There is no $j_\sp$ and $j_\ch$ inhomogenity for S constant
and nearly no inhomogenity in case of column type structure as the
$j_\sp$ and $j_\ch$ are homogenized by a Co1 layer. For constriction
type structure the inhomogenity is 12\% for $\up\up$ and 5\% for
$\up\dw$. Inhomogenity is defined as
$(j_\mathrm{max}-j_\mathrm{min})/(j_\mathrm{max}+j_\mathrm{min})$,
where maximal, minimal value is taken from cross-sectional current
distribution in Fig.~\ref{f:prftra}. The inhomogenity is reduced for
$\up\dw$ due to different conductivities of up and down channel
conductivities, leading to other current homogenization in Cu2 spacer
layer.  If Co1 layer is infinitly large, the inhomogenity is increased
to 10\% and 29\% for $\up\up$ and $\up\dw$, respectively.  The $j_\ch$
inhomogenity is 8\% and 20\% for constriction type structure with and
without infinitely large Co1 layer, respectively.

Here remains a question whether such a $j_\sp$ inhomogenities make
current magnetization reversal easier or not. Adventage of $j_\sp$
inhomogeneity may be, that it localy enhanced $j_\sp$ near the pillar
edge, wheas decreasing $j_\sp$ at the pillar center.  As shown in
Table~\ref{t:tab}, the mean value of $j_\sp$ is about the same for
(Cu/Co)$^2$ constriction structure with and without infinitely large
Co1 layer. However, in the second case, $j_\sp$ is much more
inhomogenous.  This is not well presented in Fig.~\ref{f:prftra}(a),
we do not see the largest $j_\sp$ flowing in the vicinity of the
corners of the square pillars.  Disadventage of $j_\sp$ inhomogenity
may be different magnitude of torgue exerted on magnetic spins of free
Co2 layer. This may be particulary important in the case of high speed
switching associated with magnetization precession.

\subsection{Electrochemical potential inside structure}
\label{s:mu}

\figXI

Figure~\ref{f:mu} presents profiles of $\mu_\up$, $\mu_\dw$ and $\a$
(hereafter $\mu$-profiles) along the center axis of the (a)
(Cu/Co)$^2$ $\up\up$, $\up\dw$ (b) (Cu/Co)$^3$ $\up\up\up$,
$\up\dw\up$ and (c) (Cu/Co)$^3$ $\up\up\dw$, $\up\dw\dw$. S-constant
structure is presented for both magnetization directions of Co2
layers, although column and constriction types are presented only for
$\up$ magnetization of Co2 layer.  In contrast to $j_\sp$,
inhomogenity of spin accumulation $\Delta\mu$ at the position of free
Co2 layer is very small, mostly bellow 1\%. Table~\ref{t:tab} presents
a~mean value of $\Delta\mu$ for all types of the studied structures.

Figures~\ref{f:mu} (a)(b) and (c) show that $\mu$-profiles depend 
slightly on magnetization of free Co2 layer because
$t_\mathrm{Co2}\ll \lambda_\co$. Figure~\ref{f:mu}(a) exhibits
suppression of $\Delta\mu$ in the vicinity of infinitely large Cu layer.
The reason is exactly the same as discussed in
Sec.~\ref{s:cr-through}: the infinitely large Cu layer works as a strong
spin-scatterer, causing a~small spin-flip resistance (large scattering)
between up and down channels. Obviously, such a shortcut reduces
$\Delta\mu$.

This is contradictory to \cite{ura03a,ura03b,jia04}, where it is argued
that presence of spin-scatterer increases spin acululation $\Delta\mu$
inside pillar. It should be emphesised that presence of spin-scatterers
increases $j_\sp$ (and magnetoresistance) in the pillar, but
\textit{reduce} $\Delta\mu$.

Table~\ref{t:tab} shows that the largest $\Delta\mu$ is obtained for
column type structure, by 20\% larger than for S-constant type
structure. The reason is as follow: when Cu1 is not infinitely large
(S constant structure), $\Delta\mu$ changes its sign approximately in
the middle of Co1 layer [Fig.~\ref{f:mu}(a) S constant]. When Cu1 is
infinitely large, it acts as strong spin-scatterer and shortcuts up
and down channels. Hence, $\Delta\mu$ at Cu1/Co1 interface is nearly
zero and hence larger $\Delta\mu$ is obtained at the Co1/Cu2 interface
and inside Co2 layer [Fig.~\ref{f:mu}(a) column]. But to realize this,
it is necessary that up and down channels above free Co2 layer should
not be shortcut meaning that the cover layer has not to be infinite.

Figure~\ref{f:mu}(b) shows that $\Delta\mu$ is reduced when Co1 and
Co3 layers have parallel magnetization configuration.
Figure~\ref{f:mu}(c) shows an increase in $\Delta\mu$ when Co1 and Co3
layers are antiparallel, enhancing $\Delta\mu$ by
factor of~2 with respect to (Cu/Co)$^2$ S constant structure. In this
case the types of structure are not so important as in (Cu/Co)$^2$
case, because $\Delta\mu$ at the position of Co2 free layer is
``screened'' by spin-scattering inside Co1 and Co3 layers.

Hence, there is effectively no spin-accumulation $\Delta\mu$ in the case of
a~commonly used constriction type structure with two FM layers. It
may explain why this contribution to magnetization reversal (predicted
in \cite{hei01}) was not observed in Co/Cu structure \cite{alb02}.  To
obtain non-zero $\Delta\mu$, it is necessary to use either 3 FM layer
system with antiparallel configuration of first and last FM layers, or
to ensure that structure above free layer will not contain
spin-scatterers. It can be reached when pillar structure above free
layer (i) does not contain any strong spin-scatterer layers (as Au, Ag,
Pt, Ru) and (ii) the cover layer does not contain large volume of
metal. It means that pillar current drain should be realized by a long
pillar or by a thin cover layer.

\subsection{Magnetoresistance}
\label{s:mr}

\figXII

Finally, we discuss influence of type structure on magnetoresistance ratio
(MR), presented in the last column in Table~\ref{t:tab}. The value of
MR is determined as
MR$=(\Delta\a_{\up}-\Delta\a_{\dw})/(\Delta\a_{\up}+\Delta\a_{\dw})$,
where $\up$, $\dw$ denotes magnetization of free Co2 layer,
$\Delta\a_{\updw}=\a_{\mathrm{last},\updw}-\a_{\mathrm{first},\updw}$,
where $\a_{\mathrm{first},\updw}$ and $\a_{\mathrm{last},\updw}$ are
determined on Cu side of the first and last Cu/Co interface,
respectively.

The calculated value of MR in the case of (Cu/Co)$^2$ S-constant
structure is 0.48\%. However, in the case of (Cu/Co)$^2$ constriction
type, it reachs 1.01\% (enhancement by a~factor of~2) and in the case
of the constriction with infinite Co1 layer even 1.43\% (enhancement
by a~factor of~3). The last value may be misleading as this increase
is mainly due to resistivity reduction of Co1 infinite layer. 

Table~\ref{t:tab} shows MR is strongly affected by the type structure,
increasing with an increase of $j_\sp$ rather than $\Delta\mu$. This
tendency may be observed also for (Co/Cu)$^3$ layer, showing increase
(decrease) in MR when Co1 and Co3 layers have parallel (antiparallel)
magnetization configuration.

Figure~\ref{f:jR}(a) shows dependance of $j_\sp$ on $A\Delta R$, where
$\Delta R=R_{\up}-R_\dw$, where $\up$, $\dw$ means up, down
magnetization of Co2 layer, respectively. It shows that $j_\sp$ is
proportional to $\Delta R$.  This figure is analogous to the
experimental dependence of critical switching current $J_{s,\ch}$ as a
function of $\Delta R$ \cite{ura03a,ura03b,jia04}, where they found that
$1/J_{s,\ch}$ is proportional to $\Delta R$.

These graphs are analogous from following reasons: to switch magnetic
layer we need to overcome the critical spin-polarized current density
\begin{equation}
j_{s,\sp,0}=p J_{s,\ch}/S_\mathrm{pillar},
\label{eq:jspcr}
\end{equation}
where $J_{s,\ch}$ is critical switching charge current and $p$ is
spin-polarized current injection efficiency. Because $j_{s,\sp,0}$ and
$S_\mathrm{pillar}$ are constant, so $p\sim 1/J_{s,\ch}$.  On the other
hand, from our calculations, we can express $p$ as
$p=j_\sp/(J_\ch/S_\mathrm{pillar})$, where $J_{\ch}$ is a fixed charge
current flowing to the structure; therefore $p\sim j_\sp$.  However,
in our model, $j_\sp$ is slightly larger for Co1 and Co2 parallel
configuration (Fig.~\ref{f:crsp}), although in \cite{ura03a,ura03b,jia04},
$1/J_{s,\ch}$ is larger for antiparallel configuration, in agreement
with the spin-transfer model \cite{slo96}.  Figure~\ref{f:jR}(a) also
provides an~important consequence: when changing electric properties of
surroundings of FM(fixed)/spacer/FM(free) layers, an enhancement of
$j_\sp$ in the position of free layer leads to enhancement of MR.

Figure~\ref{f:jR}(b) shows dependence of $\Delta\mu$ on $\Delta R$. We
can see that with increasing $\Delta R$, $\Delta\mu$ is reduced.  Two
different slopes correspond to two different ``sources'' $\Delta\mu$
acting as with different ``hardness''.  The source hardness is
determined by a presence or absence of scatterers bellow fixed Co1
layer. Hard source is for constriction and column structure type, i.e.
when Cu1 is infinitely large. Weak one is for S constant structure
type, i.e. when for Cu1 is not infinitely large.  The explanation of
this behavior has been already provided in previous
Section~\ref{s:mu}: when spin-flip scatterer is presented bellow fixed
Co1 layer, it vanishes $\Delta\mu$ on Cu1/Co1 interface and hence
provides harder source of $\Delta\mu$.  Figure~\ref{f:jR} also shows
that when changing surroundings of FM(fixed)/spacer/FM(free) layers,
an increase in $\Delta\mu$ at the position of free layer is related to
a~decrease of MR.

Remark, that for constriction type structure with Co1 layer infinitely
large [case(d) in Fig.~\ref{f:types}], the ``source'' of $\Delta\mu$
becomes softer than for constriction type [case (c) in
Fig.~\ref{f:types}].


\section{Expression of infinitely large layers by 1D models}
\label{s:cr1dm}

As we have shown, the constriction type structures modify the
profile of both $j_\sp$ and $\mu_{\up,\dw}$ with respect to S constant
types, which provides equivalent results to 1D VF formalism
\cite{val93}.  However, to describe $j_\sp$ and $\mu_\updw$ inside
structure by 3D model may be tedious procedure. That is why here we discuss
briefly, how to modified VF model to take into account infinitely large
layers.

We propose, that each layer can have its own cross-sectional area
$S_i$.  Then, instead of conserving spin polarized current density
$j_\updw$ in the 1D formalism, we propose to conserve spin polarized
current $J_\updw$.  Actually, we have used this boundary condition
already in Section~\ref{s:multilayer}. This boundary condition is
justified when (i) thickness of the layer $L_i$ is thick enough
compared to the change of pillar diameter between neighboring layers,
$|a_{i+1}-a_{i}|<L_i$, $|a_{i}-a_{i-1}|<L_i$, where
$a_i\approx\sqrt{S_i}$, so that current has enough space to spread to
different cross-sectional area $S_i$ (ii) pillar diameter $a_i$ is
smaller than spin-flip length $\lambda$. It can be shown that if
$S_i\rightarrow\inf$, then $j_\updw$, $\mu_\updw$, MR etc.\ converge.
So, it is not important, how large value $S_i$ is used to describe
infinitely large layers.

Figure~\ref{f:1dm} shows $j_\sp$ and $\mu_\updw$ calculated by 3D
models in the case of constriction type structure. 
These data are compared with 1D models with and without expanding the
terminating Cu layers, called ``1D constriction''
and ``S constant'', respectively. 

3D constriction is described by 1D constriction (S constant) model
with precision of 15\% (50\%) for $j_\sp$ and 40\% (80\%) for
$\Delta\mu$ [Fig.~\ref{f:1dm}(a)(b)]. The agreement between
$\Delta\mu$ for constriction type structure is quite poor, because
$\Delta\mu$ at this configuration is very small. In the case of
$\Delta\mu$ larger than the above case,
the agreement is about 10\%. 

This 1D model has been used to calculate $j_\sp$, $\Delta\mu$ and MR
in the position of free Co2 layer for all types of studied structures.
The results are presented in Table~\ref{t:tab} in square brackets; we
can see rather good agreement in all cases.  An exception is the case
of constriction type structure with infinitely large Co1 layer, as in
this case the condition (ii) is not fulfilled.

\section{Conclusion}
\label{s:con}

We have developed formalism which allows to calculate spin-polarized
current $j_\sp$, and electrochemical potential $\mu_\updw$ inside
arbitrary electric circuit, consisting of ferro- or
non-magnetic metallic SDRE elements as well as
interface and surface resistivities. The formalism is limited
to the parallel/antiparallel magnetic orientation in diffusive regime.

To calculate spatial distribution of $\mu_\updw$, $j_\updw$ inside
nanostructure, we divide the structure into an 1D, 2D or 3D
electric circuit network of SDRE elements which is successively solved. When
division is carried out as described in Section~\ref{s:grid}, the
renormalization of spin-flip length $\lambda$ has to be performed.

This formalism is applied to (Cu/Co)$^2$ and (Cu/Co)$^3$, 
pillar structures, where pillar cross-sectional area of
starting/terminating layers were assumed to be either infinitely large
(column type, constriction type) or they have the same cross-sectional
area as pillar (S const types).

Inside the pillar surrounded by infinitely large layers, the $j_\sp$,
$j_\ch$ can be inhomogeneous. Maximal inhomogeneity is found to be 29\%
and 20\% in the case of infinitely large Co1 layer. On the other hand,
profile of $\Delta\mu$ is found more homogeneous, with found maximum 
inhomogeneity of 2\%. Such $j_\sp$ inhomogeneities may locally enhance
the value of $j_\sp$, but they may disturb the magnetization
reversal associated with spin precession. Due to
$\sigma_\co\ll\sigma_\cu$, $AR_{\lambda,\co}>>AR_{\lambda,\cu}$, both
$j_\ch$ and $j_\sp$ flow rather perpendicular to the Co layers and
furthermore the presence of Co layer makes $j_\sp$, $j_\ch$ more
homogeneous.

When pillar is terminated by infinitely large layers, they serve
as spin-scatterers, shortcutting up and down channels and hence
modifying profiles of $\Delta\mu$ and $j_\sp$.  

When such a spin-scatterers (but it is also valid for different
spin-scatterers, as layers of Au, Ag, Pt, Ru, etc.) are introduced
bellow ``fixed'' Co1 layer, they make $\Delta\mu$ source ``harder''.
When they are placed above free Co2 layer, they 
shortcut up and down channels, reducing $\Delta\mu$ nearly
to zero, and hence enhancing $j_\sp$.

Consequently, to get maximum $j_\sp$ in the case of (Cu/Co)$^2$ at the
position of ``free'' Co2 layer, it is important to introduce
spin-scatterers both bellow fixed Co1 layer and above free Co2
layer.  To get maximum $\Delta\mu$, it is important to introduce
spin-scatterer bellow fixed Co1 layer, but reduce spin-scattering above
Co2 free layer, latter one can be realized by reducing volume of material
above Co2 layer, for example current drain can be thin long
nanowire or cover layer with very thin thickness.  As most of
experimentally studied (Cu/Co)$^2$ structures have infinitely large
layers at both ends, $\Delta\mu$ inside them is nearly zero.

For (Cu/Co)$^3$ system, maximum $j_\sp$ ($\Delta\mu$) is for parallel
(antiparallel) magnetization of first, last Co layer. When applying
above described optimizations, $j_\sp$ and $\Delta\mu$ can be further
enhanced.

Furthermore, in agreement with experimental results \cite{ura03a,ura03b,jia04}
we found that $j_\sp$ is linearly proportional to $\Delta
R=R_{\dw}-R_\up$ (and hence MR is increased when increasing $j_\sp$ at the
position of free Co2 layer).  Furthermore, dependence of
$\Delta\mu(\Delta R)$ is also linear, but $\Delta\mu$ is reduced when
increasing $\Delta R$ (when increasing MR).

Finally, we propose simple modification of 1D Valet-Fert
formalism \cite{val93}, to incorporate spin-scattering induced by
infinitely large layers attached to pillar.



\end{document}